\documentclass[a4paper,12pt]{article}

\usepackage{amsmath,amssymb,algorithmic}
\usepackage{latexsym}
\usepackage[latin1]{inputenc}
\usepackage[T1]{fontenc}
\usepackage{textcomp}
\usepackage{times}

\usepackage[pdftex]{graphicx}
\usepackage[frenchb]{babel}
\usepackage{babelbib}
\usepackage{hyperref}
\hypersetup{
colorlinks=true, %colorise les liens
citecolor=blue, %
pdftoolbar=false,
bookmarksopen=true,
breaklinks=true, %permet le retour à la ligne dans les liens trop longs
urlcolor=blue, %couleur des hyperliens 
linkcolor=blue, %couleur des liens internes
}

\newcommand{\bm}[1]{\mathbf{ #1}}
\newcommand{\bs}[1]{\boldsymbol{#1}}

\author{Philippe Besse\thanks{Université de Toulouse -- INSA, Institut de Mathématiques, UMR CNRS 5219} \and Nathalie Villa-Vialaneix\thanks{INRA -- UR875 MIA-T}}
\title{Statistique et \emph{Big Data Analytics} \\ Volumétrie -- L'Attaque~des~\emph{Clones}}

\begin{document}
\sloppy
\DeclareGraphicsExtensions{.pdf,.jpg}
%\graphicspath{{D:/Users/pbesse/Dropbox/Wikistat/M2-BigData/Figures/}{../Figures}}
\maketitle

\begin{quote}
{\bf Résumé}: Cet article suppose acquises les compétences et expertises d'un
statisticien en apprentissage non supervisé
(\href{http://wikistat.fr/pdf/st-m-explo-nmf.pdf}{NMF},
\href{http://wikistat.fr/pdf/st-m-explo-classif.pdf}{$k$-means},
\href{http://wikistat.fr/pdf/st-m-explo-alglin.pdf}{svd}) et supervisé
(\href{http://wikistat.fr/pdf/st-m-modlin-regmult.pdf}{régression},
\href{http://wikistat.fr/pdf/st-m-app-cart.pdf}{cart},
\href{http://wikistat.fr/pdf/st-m-app-agreg.pdf}{random forest}). Quelles
compétences et savoir faire ce statisticien doit-il acquérir pour passer à
l'échelle ``\emph{Volume}'' des grandes masses de données ? Après un rapide tour
d'horizon des différentes stratégies offertes et principalement celles imposées
par l'environnement \emph{Hadoop}, les algorithmes des quelques méthodes
d'apprentissage disponibles sont brièvement décrits pour comprendre comment ils
sont adaptés aux contraintes fortes des fonctionnalités \emph{Map-Reduce}. La prochaine étape sera sans doute de les réécrire dans le langage matriciel similaire à R des communautés \emph{Mahout}, \emph{Spark}, \emph{Scala}. 

{\bf Mots-clefs}: Statistique; Fouille de Données; Grande Dimension; Apprentissage Statistique; Datamasse; algorithmes; Hadoop, Map-Reduce; Scalability.

{\bf Abstract}: 
This article assumes acquired the skills and expertise of a statistician in unsupervised  (NMF, k-means, SVD) and supervised learning (regression, CART, random forest). What skills and knowledge do the statistician must acquire it to reach the "Volume" scale of big data? After a quick overview of the different strategies available and especially those imposed by Hadoop, algorithms of some available learning methods  are outlined to understand how they are adapted to high stresses of  Map-Reduce functionalities. The next step will probably be to rewrite them using the R like matricial language which is developped by the communities\emph{Mahout}, \emph{Spark} and \emph{Scala}.

\end{quote}

% un input pour mettre en forme le fichier pour Hal ou pour wikistat
\section{Introduction}
\subsection{Motivations}
L'historique récent du traitement des données est schématiquement relaté à
travers une odyssée:
\href{http://hal.archives-ouvertes.fr/hal-00959267}{\emph{Retour vers le Futur
III}} (Besse et al.; 2014)\cite{besgl14} montrant comment un statisticien est
successivement devenu prospecteur de données, bio-informaticien et maintenant
\emph{data scientist}, à chaque fois que le volume qu'il avait à traiter était
multiplié par un facteur mille. Cette présentation rappelle également les trois
aspects : volume, variété, vélocité, qui définissent généralement le \emph{Big
Data} au sein d'un \emph{écosystème}, ou plutôt une jungle, excessivement
complexe, concurrentielle voire, conflictuelle, dans laquelle le mathématicien /
statisticien peine à se retrouver; ce sont ces difficultés de béotiens qui
motivent la présentation d'une nouvelle saga: \emph{Data War} en 5 volets dont
seulement trois seront interprétés.

\begin{figure}
\centerline{\includegraphics[width=6cm]{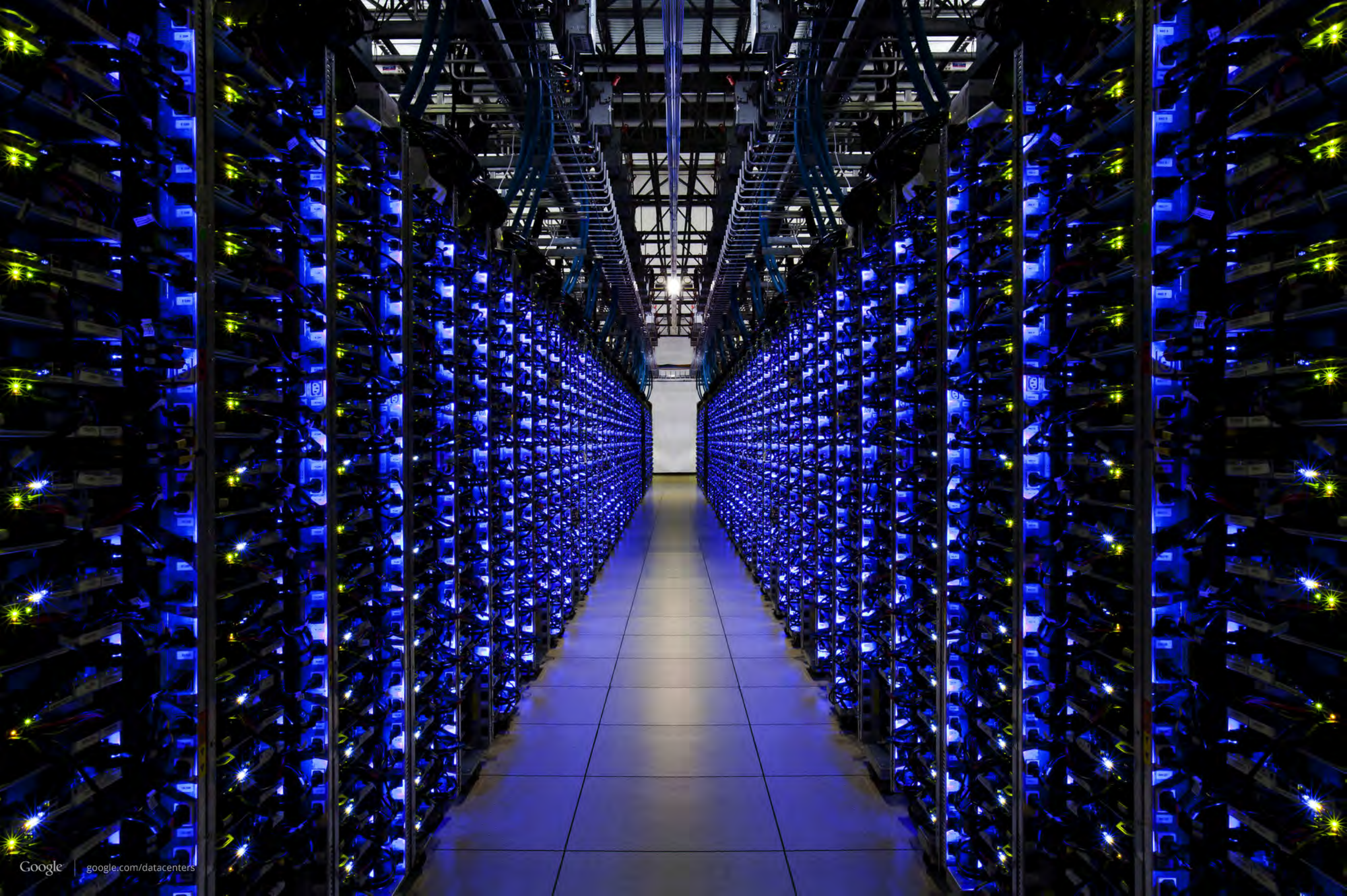}}
\caption{\it Armée de \emph{clones}, baies de serveurs, alignés par milliers dans le hangar d'un \href{http://www.google.com/about/datacenters/}{centre de données} de Google.}\label{datacenter}
\end{figure}

\begin{description}
	\item[La Menace Fantôme] est jouée avec la NSA dans le rôle de \emph{Big Brother},
	\item[L'Attaque des Clones] contre la \emph{Volumétrie} est l'objet du présent article avec les baies ou conteneurs d'empilements de serveurs dans le rôle des clones (cf. figure \ref{datacenter}).
	\item[La Revanche des Maths] intervient dans la \href{http://wikistat/pdf/st-stat-bigdata-variete.pdf}{vignette} (à venir) consacrée à la \emph{Variété} (courbes, signaux, images, trajectoires, chemins...) des données qui rend indispensable des compétences mathématiques en modélisation, notamment pour des données industrielles, de santé publique...
	\item[Le Nouvel Espoir] est d'aider au transfert des technologies et méthodes efficaces issues du e-commerce vers d'autres domaines (santé, climat, énergie...) et à d'autres buts que ceux  mercantiles des origines.
	\item[L'Empire Contre-Attaque] avec GAFA\footnote{Google, Apple, Facebook, Amazon} dans le rôle principal de la \href{http://wikistat/pdf/st-stat-bigdata-velocite.pdf}{vignette} (à venir) qui aborde les problèmes de flux ou \emph{Vélocité} des données.
\end{description}
\centerline{\includegraphics[width=3cm]{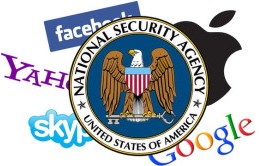}}

\subsection{Environnement informatique}
Avant d'aborder le c\oe ur du sujet, l'analyse à proprement parler, de données
massives, nous introduisons en décrivant brièvement le panorama de la manière dont les données
sont stockées et gérées d'un point de vue informatique. En effet, le contexte
technique est étroitement lié aux problèmes posés par l'analyse de données
massives et donc indispensable pour en appréhender les enjeux. Nous faisons,
dans cette présentation, une place particulière à \emph{Hadoop}, qui est l'un des
standards informatiques de la gestion de données massives sur lequel nous avons
pris le parti de nous appuyer pour illustrer cet article.

Le contexte général de cet article est celui de la \emph{distribution} des
ressources informatiques. On parle de système distribué dans le cadre où une
collection de postes (ordinateurs ou n\oe uds) autonomes sont connectés par un
réseau. Ce type de fonctionnement permet de partager les capacités de chacun des
postes, capacités qui se déclinent en :
\begin{itemize}
	\item partage des ressources de calcul ou \emph{calcul distribué},
\emph{partagé} ou \emph{parallèle}. Il s'agit ici de répartir un calcul (une
analyse de données au sens statistique du terme, par exemple) sur un ensemble
d'ententités de calcul indépendantes (\emph{c\oe urs} d'un micro-processeur par
exemple). Ces entités peuvent être situés sur un même ordinateur ou, dans le
cas qui nous préoccupe dans cet article, sur plusieurs ordinateurs fonctionnant
ensemble au sein d'un \emph{cluster de calcul}. Cette dernière solution permet
d'augmenter les capacités de calcul à moindre coût comparé à l'investissement
que représente l'achat d'un super-calculateur ;
	\item partage des \emph{ressources de stockage des données}. Dans le cadre
d'un calcul distribué, on distingue deux celui où la mémoire (où sont stockées
les données à analyser) est accédée de la même manière par tous les processeurs 
(mémoire partagée) et celui où tous les processeurs n'accèdent pas de la même 
manière à toutes les données (mémoire distribuée). 
\end{itemize}

Dans les deux cas, des environnements logiciels (c'est-à-dire, des ensemble de
programmes) ont été mis en place pour permettre à l'utilisateur de gérer le
calcul et la mémoire comme si il n'avait à faire qu'à un seul ordinateur : le
programme se charge de répartir les calculs et les données. \emph{Hadoop} est un de ces
environnements : c'est un ensemble de programmes (que nous décrirons plus
précisément plus tard dans cet article) programmés en java, distribués sous
licence libre et destinés à faciliter le déploiement de tâches sur plusieurs
milliers de n\oe uds. En particulier, \emph{Hadoop} permet de gérer un système de
fichiers de données distribués : HDFS (Hadoop Distributed File System) permet de
gérer le stockage de très gros volumes de données sur un très grand nombre de
machines. Du point de vue de l'utilisateur, la couche d'abstraction fournie par
HDFS permet de manipuler les données comme si elles étaient stockées sur un
unique ordinateur. Contrairement aux bases de données classiques, HDFS prend en
charge des \emph{données non structurées}, c'est-à-dire, des données qui se
présentent sous la forme de textes simples comme des fichiers de log par
exemples.

	L'avantage de HDFS est que le système gère la localisation des données lors de
la répartition des tâches : un n\oe ud donné recevra la tâche de traiter les
données qu'il contient afin de limiter le temps de transfert de données qui peut
être rapidement prohibitif lorsque l'on travaille sur un très grand nombre de
n\oe uds. Pour ce faire, une approche classique est d'utiliser une
décomposition des opérations de calcul en étapes \og Map\fg\ et \og Reduce\fg\
(les détails de l'approche sont donnés plus loin dans le document) qui
décomposent les étapes d'un calcul de manière à faire traiter à chaque n\oe ud
une petite partie des données indépendamment des traitements effectuées sur les
autres n\oe uds.

Une fois les données stockées, ou  leur flux organisé, leur \emph{valorisation}
nécessite une phase d'analyse (\emph{Analytics}). L'objectif est de tenter de
pénétrer cette jungle pour en comprendre les enjeux, y tracer des sentiers
suivant les options possibles afin d'aider à y faire les bons choix. Le parti
est pris d'utiliser, si possible au mieux, les ressources logicielles (langages,
librairies) \emph{open source} existantes en tentant de minimiser les temps de
calcul tout en évitant la programmation, souvent re-programmation, de méthodes
au code par ailleurs efficace. Donc minimiser, certes les temps de calcul, mais
également les coûts humains de développement pour réaliser des premiers
prototypes d'analyse.

La littérature électronique sur le sujet est, elle aussi, massive (même si 
assez redondante) et elle fait émerger un ensemble de :

\subsection{Questions}

La première part du constat que, beaucoup des exemples traités, notamment 
ceux des besoins initiaux, se résument principalement à des dénombrements, 
d'occurrences de mots, d'événements
(décès, retards...) de co-occurences (règles d'association). Pour ces objectifs,
les architectures et algorithmes parallélisés avec \emph{MapReduce} sont
efficaces lorsqu'ils traitent l'ensemble des données même si celles-ci sont très
volumineuses. Aussi, pour d'autres objectifs, on s'interroge sur la réelle 
nécessité,
en terme de qualité de prévision d'un modèle, d'estimer celui-ci sur l'ensemble
des données plutôt que sur un échantillon représentatif stockable en mémoire.

La deuxième est de savoir à partir de quel volume un environnement spécifique 
de stockage comme
\emph{Hadoop} s'avère nécessaire. Lorsque les données peuvent être 
stockés sur un cluster de calcul ``classique'', cette solution (qui limite le 
temps d'accès aux données) est plus efficace. Toutefois, au delà 
d'un certain volume, \emph{Hadoop} est une solution pertinente pour gérer des 
données très volumineuses réparties sur un très grand nombre de machines. 
Ainsi, avant de se diriger vers une solution de type \emph{Hadoop}, il faut se 
demander si un accroissement de la mémoire physique et/ou l'utilisation d'une 
librairie (R) simulant un accroissement de la mémoire vive ne résoudrait pas le 
problème du traitement des données sans faire appel à \emph{Hadoop} : 
dans beaucoup de situations, ce choix s'avèrera plus judicieux et plus 
efficace. La bonne démarche est donc d'abord de se demander quelles sont les 
ressources matérielles et logicielles disponibles et quels sont les algorithmes 
de modélisation disponibles dans l'environnement utilisé. On se limitera au 
choix de \emph{Hadoop} lorsque celui-ci est imposé par le volume des données ou 
bien que \emph{Hadoop} est l'environnement donné \emph{a priori} comme 
contrainte à l'analyste de données.

Pour une partie de ces questions, les réponses dépendent de l'objectif. S'il 
semble pertinent en e-commerce de considérer tous les clients potentiels en 
première approche, ceci peut se discuter pas-à-pas en fonction du domaine et de 
l'objectif précis : \emph{clustering}, score, système de recommandation..., 
visé. Dans le cas d'un sous-échantillonnage des données, le volume peut devenir 
gérable par des programmes plus simples et plus efficaces qu'\emph{Hadoop}.

En conclusion, il est conseillé d'utiliser \emph{Hadoop} s'il n'est pas 
possible de
faire autrement, c'est-à-dire si : 
\begin{itemize}
\item on a pris le parti de vouloir tout traiter, sans échantillonnage,
avec impossibilité d'ajouter plus de mémoire ou si le temps de calcul devient
rédhibitoire avec de la mémoire virtuelle,
\item le volume des données est trop important ou les données ne sont pas
structurées, 
\item la méthodologie à déployer pour atteindre l'objectif se décompose en 
formulation \emph{MapReduce},
\item l'environnement \emph{Hadoop} est imposé, existe déjà  ou est facile à
faire mettre en place par les personnes compétentes,
\item un autre environnement plus performant n'a pas déjà pris le dessus.
 \end{itemize}

\subsection{Prépondérance de Hadoop?}

Le traitement de grandes masses de données impose une parallélisation des
calculs pour obtenir des résultats en temps raisonnable et ce, d'autant plus,
lors du traitement en temps réel d'un flux (\emph{streaming}) de données. Les
acteurs majeurs que sont Google, Amazon, Yahoo, Twitter... développent dans des
centres de données (\emph{data centers}) des architectures spécifiques pour
stocker à moindre coût de grande masses de données (pages web, listes de
requêtes, clients, articles à vendre, messages...) sous forme brute, sans format
ni structure relationnelle. Ils alignent, dans de grands hangars, des
empilements (conteneurs ou baies) de cartes mère standards de PC ``bon
marché''\footnote{La facture énergétique est le poste de dépense le plus
important: l'électricité consommée par un serveur sur sa durée de vie coûte plus
que le serveur lui-même. D'où l'importance que Google apporte aux question
environnementales dans sa
\href{http://www.google.com/about/datacenters/}{communication}.}, reliées par
une connexion réseau (cf. figure \ref{datacenter}). 

\centerline{\includegraphics[width=4cm]{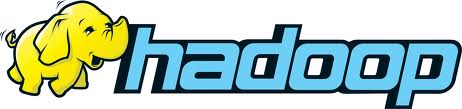}}

\emph{Hadoop} est une des solutions les plus populaires pour gérer des données
et des applications sur des milliers de n\oe uds\footnote{Un n\oe ud est une
unité de calcul, par exemple, pour simplifier, un ordinateur au sein d'un
réseau d'ordinateurs.} en utilisant des approches distribuées. \emph{Hadoop} est
composé de son propre gestionnaire de fichiers (HDFS : Hadoop Distributed File
System) qui gère des données sur plusieurs n\oe uds simultanément en permettant
l'abstraction de l'architecture physique de stockage (c'est-à-dire que
l'utilisateur n'a l'impression de travailler que sur un seul ordinateur alors
qu'il manipule des données réparties sur plusieurs n\oe uds). \emph{Hadoop} dispose
aussi d'algorithmes permettant de manipuler et d'analyser ces données de
manière parallèle, en tenant compte des contraintes spécifiques de cette
infrastructure, comme par exemple \emph{MapReduce} (Dean et Ghemawat;
2004)\cite{dea04}, HBase, ZooKeeper, Hive et Pig.

Initié par Google, \emph{Hadoop} est maintenant développé en version libre dans le
cadre de la fondation \href{http://hadoop.apache.org/}{Apache}. Le stockage
intègre également des propriétés de duplication des données afin d'assurer une
tolérance aux pannes. Lorsqu'un traitement se distribue en étapes \emph{Map} et
\emph{Reduce}, celui-ci devient ``échelonnable'' ou \emph{scalable},
c'est-à-dire que son temps de calcul est, en principe, divisé par le nombre de
n\oe uds ou serveurs qui effectuent la tâche. \emph{Hadoop} est  diffusé comme
logiciel libre et bien qu'écrit en java, tout langage de programmation peut
l'interroger et exécuter les étapes \emph{MapReduce} prédéterminées.

Comparativement à d'autres solutions de stockage plus sophistiquées: cube,
SGBDR (systèmes de gestion de base de données relationnelles), disposant d'un
langage (SQL) complexe de requêtes, et comparativement à d'autres architectures 
informatiques parallèles dans lesquelles les divers ordinateurs ou processeurs 
partagent des zones mémoires communes (mémoire partagée), \emph{Hadoop} est 
perçu, sur le plan académique, comme une régression. Cette architecture et les 
fonctionnalités très
restreintes de \emph{MapReduce} ne peuvent rivaliser avec des programmes
utilisant des langages et librairies spécifiques aux machines massivement
parallèles connectant des centaines, voire milliers, de c\oe urs sur le même
bus. Néanmoins, le poids des acteurs qui utilisent \emph{Hadoop} et le bon
compromis qu'il semble réaliser entre coûts matériels et performances en font un
système dominant pour les applications commerciales qui en découlent: les
internautes sont  des prospects à qui sont présentés des espaces publicitaires
ciblés et vendus en temps réel aux annonceurs dans un système d'enchères
automatiques.

Hormis les applications du e-commerce qui dépendent du choix préalable, souvent
\emph{Hadoop}, \emph{MongoDB}..., de l'architecture choisie pour gérer les
données, il est légitime de s'interroger sur le bon choix d'architecture (SGBD
classique ou non) de stockage et de manipulation des données en fonction des
objectifs à réaliser. Ce sera notamment le cas dans bien d'autres secteurs:
industrie (maintenance préventive en aéronautique, prospection pétrolière,
usages de pneumatiques...), transports, santé (épidémiologie), climat, énergie
(EDF, CEA...)... concernés par le stockage et le traitement de données
massives. 

\subsection{Une R-éférence}
Comme il existe de très nombreux systèmes de gestion de données, il existe de nombreux environnements logiciels \emph{open source} à l'interface utilisateur plus ou moins \emph{amicale} et acceptant ou non des fonctions écrites en  \href{http://cran.r-project.org/}{R} ou autre langage de programmation: \href{http://www.knime.org/}{KNIME}, \href{http://eric.univ-lyon2.fr/~ricco/tanagra/fr/tanagra.html}{TANAGRA}, \href{http://www.cs.waikato.ac.nz/ml/weka/}{Weka},... 

\centerline{\includegraphics[width=2cm]{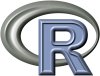}}

Néanmoins, pour tout un tas de raisons dont celle d'un large consensus au sein d'une très grande communauté d'utilisateurs, le logiciel libre \href{http://cran.r-project.org/}{R}\cite{rcor14} est une référence pour l'analyse ou l'apprentissage statistique de données conventionnelles, comme pour la recherche et le développement, la diffusion de nouvelles méthodes. Le principal problème de R (version de base) est que son exécution nécessite de charger toutes les données en mémoire vive. En conséquence, dès que le volume est important, l'exécution se bloque. Aussi, une première façon de procéder avec R, pour analyser de grandes masses, consiste simplement à extraire une table, un sous-ensemble ou plutôt un \emph{échantillon représentatif} des données avec du code java, Perl, Python, Ruby... avant de les analyser dans R. 

Une autre solution consiste à utiliser une librairie permettant une extension
virtuelle sur disque de la mémoire vive (ce qui a pour contre-partie d'alourdir
les temps d'accès aux données donc les temps de calcul). C'est le rôle des
packages {\tt ff, bigmemory}, mais qui ne gèrent pas la distribution du calcul
en parallèle, et
également aussi de ceux interfaçant R et \emph{Hadoop} qui seront plus
précisément décrits ci-dessous.

Par ailleurs, la
richesse des outils graphiques de R (\emph{i.e.} {\tt ggplot2}) permettra alors
de visualiser de façon très élaborée les résultats obtenus. Toutefois, compter
des occurrences de mots, calculer des moyennes... ne nécessitent pas la richesse
méthodologique de R qui prendrait beaucoup plus de temps pour des calculs
rudimentaires. Adler (2010)\cite{adl10} compare trois solutions pour dénombrer
le nombre de décès par sexe aux USA en 2009: 15 minutes avec RHadoop sur un
cluster de 4 serveurs, une heure avec un seul serveur, et 15 secondes, toujours
sur un seul serveur, mais avec un programme Perl. Même si le nombre restreint 
de serveur (4) de l'expérience reste réduit par rapport aux volumes habituels 
d'\emph{Hadoop}, cette expérience montre bien qu'utiliser systématiquement 
\emph{RHadoop} n'est pas pertinent.

En résumé, \emph{Hadoop} permet à R d'accéder à des gros volumes de données en
des temps raisonnables mais il faut rester conscient que, si c'est sans doute la
stratégie la plus ``simple'' à mettre en \oe uvre, ce ne sera pas la solution la
plus efficace en temps de calcul en comparaison avec d'autres environnements comme
\emph{Mahout} ou \emph{Spark} introduits ci-après et qui finalement convergent vers les fonctionnalités d'un langage matriciel identique à R.

Dans le cas contraire, les compétences nécessaires à l'utilisation des méthodes
décrites dans les parties
\href{http://Wikistat.fr/pdf/st-m-explo-intro.pdf}{Exploration} et
\href{http://Wikistat.fr/pdf/st-m-app-intro.pdf}{Apprentissage} de
\href{http://wikistat.fr}{wikistat} suffisent. Seule la forte imbrication entre
structures et volumes de données, algorithmes de calcul, méthodes de
modélisation ou apprentissage, impose de poursuivre la lecture de cet article et
l'acquisition des apprentissages qu'il développe.

Cet article se propose de décrire rapidement l'écosystème, surtout logiciel, des
grandes masses de données avant d'aborder les algorithmes et méthodes de
modélisation qui sont employés. L'objectif est d'apporter des éléments de
réponse en conclusion, notamment en terme de formation des étudiants. Les
principes des méthodes d'apprentissage ne sont pas rappelés, ils sont décrits
sur le site coopératif \href{http://wikistat.fr}{wikistat}.

\section{Environnements logiciels}
\subsection{\emph{Hadoop}}
\emph{Hadoop} est un projet \emph{open source} de la fondation \href{http://hadoop.apache.org/}{Apache} dont voici les principaux mots clefs:

\centerline{\includegraphics[width=4cm]{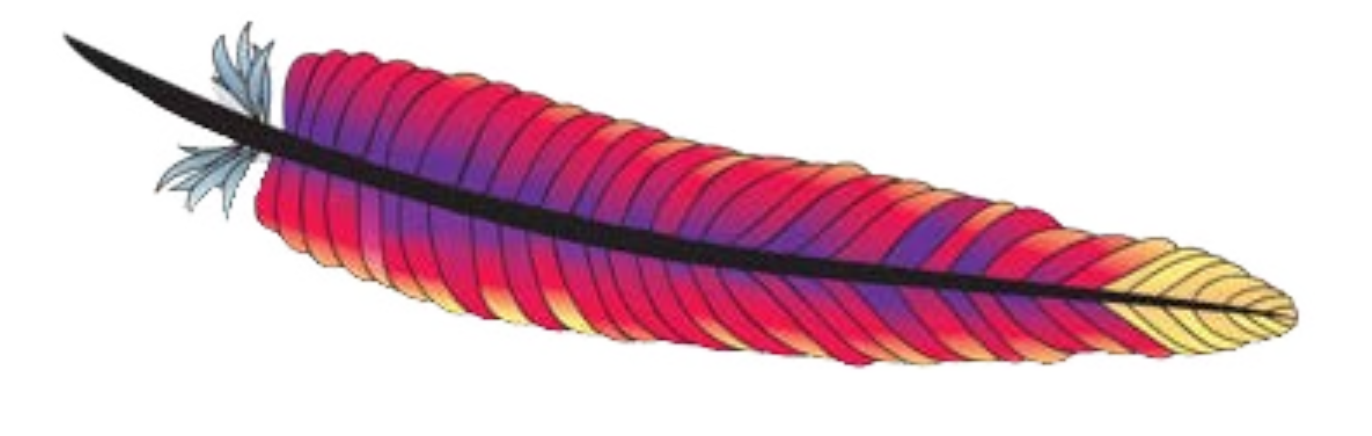}\includegraphics[width=4cm]{Hadoop}}
\begin{description}
\item[HDFS] (\emph{hadoop distributed file system}) système de fichiers
distribués sur un ensemble de disques (un disque par n\oe ud ou serveur) mais
manipulés et vus de l'utilisateur comme un seul fichier. Réplication des données
sur plusieurs hôtes pour la fiabilité. HDFS est \emph{fault tolerant}, sans
faire appel à des duplications, si un hôte ou un disque a des
problèmes.
\item[MapReduce] est un principe de programmation informatique qui est au c\oe
ur de la parallélisation des traitements dans \emph{Hadoop} et qui comprend 
deux étapes : map et reduce. Il est décrit plus précisément dans la 
section~\ref{mapreduce-section} de ce document.
\item[Hive] langage développé par Facebook pour interroger une base \emph{Hadoop} avec une syntaxe proche du SQL (\emph{hql ou Hive query language}).
\item[Hbase] base de données orientée ``colonne'' et utilisant HDFS.  
\item[Pig] langage développé par Yahoo similaire dans sa syntaxe à Perl et visant les objectifs de \emph{Hive} en utilisant le langage \emph{pig latin}.
\item[Sqoop] pour (Sq)l to Had(oop) permet le transfert entre bases SQL et \emph{Hadoop}.
\end{description}
Plusieurs distributions \emph{Hadoop} sont proposées :
\href{http://hadoop.apache.org/}{Apache},
\href{http://blog.cloudera.com/}{Cloudera},
\href{http://hortonworks.com/}{Hortonworks} à installer de préférence sous
Linux (des solutions existent toutefois pour installer \emph{Hadoop} sous Windows). Ces
distributions proposent également des machines virtuelles pour faire du
\emph{Hadoop} avec un nombre de n\oe uds très limité (un seul par exemple) et
donc au moins tester les scripts en local avant de les lancer en vraie grandeur.
La mise en place d'un tel environnement nécessite des compétences qui ne sont
pas du tout abordées. Pour un essai sur un système distribué réel, il est 
possible d'utiliser les services d'AWS (\href{http://aws.amazon.com}{Amazon Web 
Services}) : c'est souvent
cette solution qui est utilisée par les \emph{start-up} qui prolifèrent dans le
e-commerce mais elle ne peut être retenue par une entreprise industrielle pour
des contraintes évidentes de confidentialité, contraintes et suspicions qui
freinent largement le déploiement de \emph{clouds}.

\subsection{\emph{Hadoop streaming}}
Bien qu'\emph{hadoop} soit écrit en java, il n'est pas imposé d'utiliser ce
langage pour analyser les données stockées dans une architecture \emph{Hadoop}.
Tout langage de programmation (Python, C, java... et même R) manipulant des
chaînes de caractères peut servir à écrire les étapes \emph{MapReduce} pour
enchaîner les traitements. Bien que sans doute plus efficace, l'option de
traitement en \emph{streaming} n'est pour l'instant pas développée dans cet
article pour favoriser celle nécessitant moins de compétences ou de temps de
programmation.

\subsection{Mahout}
\href{http://mahout.apache.org/}{\emph{Mahout}} (Owen et al. 
2011)\cite{oweadf11} est également un projet de la fondation Apache. C'est une 
Collection de méthodes programmées en java et destinées au machine learning sur 
des architectures avec mémoire distribuée, comme Hadoop. Depuis avril 2014, 
Mahout n'inclut plus dans ses nouveaux développements d'algorithmes conçus pour 
la méthodologie de programmation \emph{Hadoop - MapReduce}\footnote{Voir 
\url{https://mahout.apache.org} : ``\emph{The Mahout community decided to move 
its codebase onto modern data processing systems that offer a richer programming 
model and more efficient execution than Hadoop MapReduce. Mahout will therefore 
reject new MapReduce algorithm implementations from now on. We will however keep 
our widely used MapReduce algorithms in the codebase and maintain them.}''}

\centerline{\includegraphics[width=4cm]{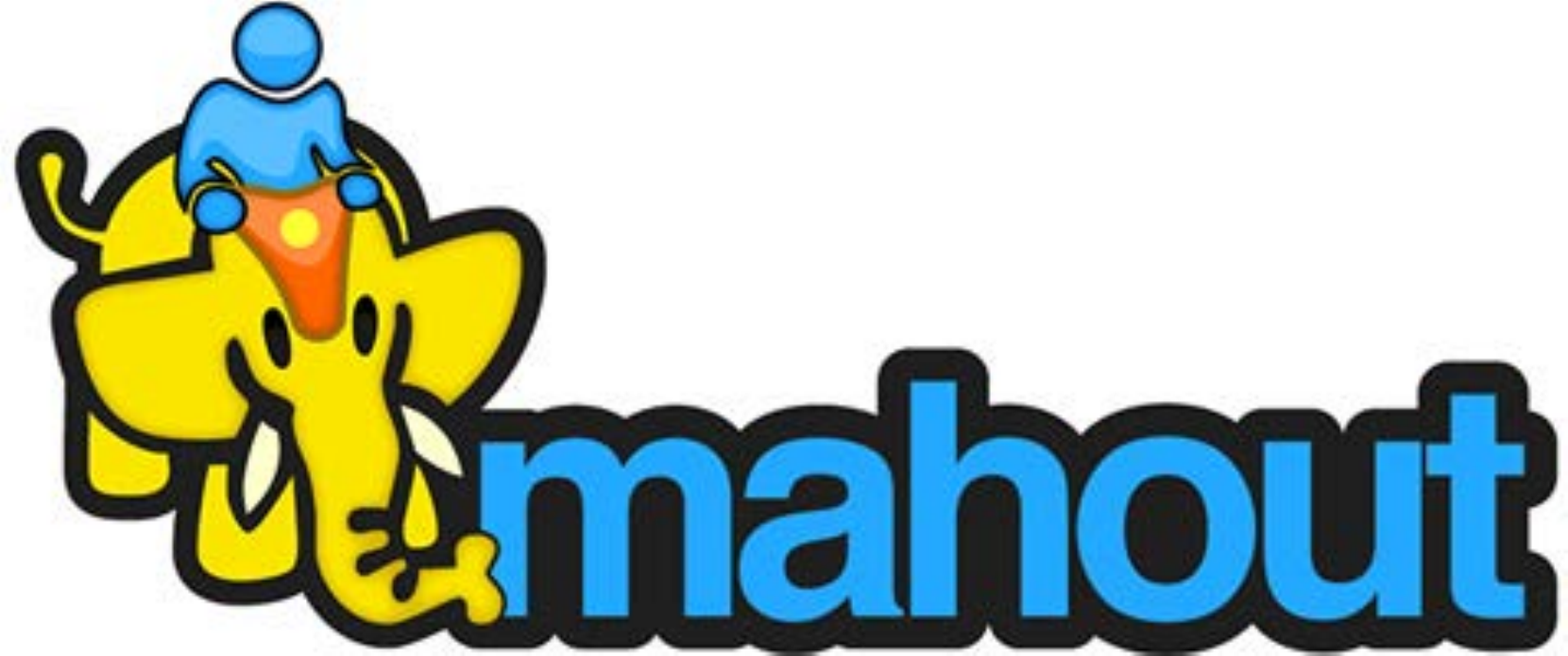}}

Mahout intègre des programmes pour la recommandation utilisateur, la NMF, la 
classification non supervisée par $k$-means, la régression logistique, les 
forêts aléatoires... La communauté de développeurs se décrit comme dynamique 
mais des compétences en Java sont indispensables pour utiliser les programmes. 

Toutefois, les développements en cours avec l'abandon de \emph{Hadoop} et le rapprochement avec \emph{Spark} ci-dessous autour d'un langage matriciel similaire à R ouvre de nouveaux horizons.

\subsection{Spark}
L'architecture de fichiers distribuée de \emph{Hadoop}
est particulièrement adaptée au stockage et à la parallélisation de tâches
élémentaires. La section suivante, sur les algorithmes, illustre les fortes
contraintes imposées par les fonctionnalités \emph{MapReduce} pour le
déploiement de méthodes d'apprentissage. Dès qu'un algorithme est itératif,
(\emph{i.e.} $k$-means), chaque itération communique avec la suivante par une
écriture puis lecture dans la base HDFS. C'est extrêmement contraignant et
pénalisant pour les temps d'exécution, car c'est la seule façon de faire
communiquer les n\oe uds d'un cluster en dehors des fonctions \emph{MapReduce}.
Les temps d'exécution sont généralement et principalement impactés par la
complexité d'un algorithme et le temps de calcul. Avec l'architecture
\emph{Hadoop} et un algorithme itératif, ce sont les temps de lecture et
écriture qui deviennent très pénalisants. Plusieurs solutions, à
l'état de prototypes (\href{http://code.google.com/p/haloop/}{HaLoop},
\href{http://www.iterativemapreduce.org/}{Twister}...), ont été proposées pour
résoudre ce problème. \emph{Spark} semble la plus prometteuse et la plus
soutenue par une communauté active. 

\centerline{\includegraphics[width=3cm]{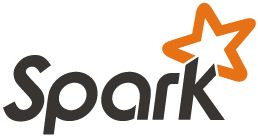}}

\href{http://spark.apache.org/}{\emph{Spark}} est devenu un projet \emph{open
source} de la fondation \href{http://hadoop.apache.org/}{Apache} dans la
continuité des travaux du laboratoire
\href{https://amplab.cs.berkeley.edu/}{amplab} de l'Université Berkley.
L'objectif est simple mais son application plus complexe surtout s'il s'agit de
préserver les propriétés de tolérance aux pannes. Il s'agit de garder en mémoire
les données entre deux itérations des étapes \emph{MapReduce}. Ceci est fait
selon un principe abstrait de mémoire distribuée: \emph{Resilient Distributed
Datasets} (Zaharia et al. 2012)\cite{zahc12}. Cet environnement est accessible
en java, \href{http://www.scala-lang.org/}{Scala}\footnote{Langage de
programmation développé à l'EPFL (Lausanne) et exécutable dans un environnement
java. C'est un langage fonctionnel adapté à la programmation d'application
parallélisable.}, Python et bientôt R (librairie SparkR). Il est accompagné
d'outils de requêtage (Shark), d'analyse de graphes (GraphX) et 
d'une bibliothèque en développement (MLbase) de méthodes d'apprentissage. 

L'objectif affiché\footnote{https://mahout.apache.org/users/sparkbindings/home.html}  par les communautés \emph{Mahout} et \emph{Spark} de la fondation Apache est l'utilisation généralisée d'un langage matriciel de syntaxe identique à celle de R. Ce langage peut manipuler(algèbre linéaire) des objets (scalaires, vecteurs, matrices) en mémoire ainsi que des matrices dont les lignes sont distribuées sans que l'utilisateur ait à s'en préoccuper. Un optimisateur d'expression se charge de sélectionner les opérateurs physiques à mettre en \oe uvre pour, par exemple, multiplier des matrices distribuées ou non. 

\subsection{Librairies R}
\subsubsection*{R parallèle}
Le traitement de données massives oblige à la parallélisation des calculs et de
nombreuses solutions sont offertes dont celles, sans doute les plus efficaces,
utilisant des librairies ou extensions spécifiques des langages les plus
courants (Fortran, java, C...) ou encore la puissance du GPU (Graphics
Processing Unit) d'un simple PC. Néanmoins les coûts humains d'écriture, mise au
point des programmes d'interface et d'analyse sont à prendre en compte. Aussi,
de nombreuses librairies ont été développées pour permettre d'utiliser au mieux,
à partir de R, les capacités d'une architecture (cluster de machines, machine
multi-c\oe urs, GPU) ou d'un environnement comme \emph{Hadoop}. Ces
librairies sont listées et décrites sur la page dédiée au
\href{http://cran.r-project.org/web/views/HighPerformanceComputing.html}{
High-Performance and Parallel Computing with R}\cite{rcor14} du CRAN
(\href{http://cran.r-project.org/}{Comprehensive R Archive Network}). La
consultation régulière de cette page est 
indispensable car, l'\emph{écosystème} R est très mouvant, voire volatile; des librairies plus maintenues disparaissent à l'occasion d'une mise à jour de R (elles sont fréquentes) tandis que d'autres peuvent devenir dominantes. 

Le parti est ici pris de s'intéresser plus particulièrement aux librairies
dédiées \emph{simultanément} à la parallélisation des calculs et aux données
massives. Celles permettant de gérer des données hors mémoire vive comme {\tt 
bigmemory} d'une part ou celles spécifiques de parallélisation comme {\tt
parallel, snow}... d'autre part, (cf. McCallum et Weston; 2011 )\cite{mccawl11}
sont volontairement laissées de côté.

\subsubsection*{R \& \emph{Hadoop}}
Prajapati (2013)\cite{pra13} décrit différentes approches d'analyse d'une base \emph{Hadoop} à partir de R.    

\begin{description}
	\item[\href{https://github.com/saptarshiguha/RHIPE/}{\tt Rhipe}] développée à l'université Purdue n'est toujours pas une librairie officielle de R et son développement semble en pause depuis fin 2012.
	\item[\href{http://cran.r-project.org/web/packages/HadoopStreaming/index.html}{\tt HadoopStreaming}] permet comme son nom l'indique de faire du \emph{streaming hadoop} à partir de R; les programmes sont plus complexes et, même si plus efficace, cette solution est momentanément laissée de côté.
	\item[\href{http://cran.r-project.org/web/packages/hive/index.html}{\tt hive}] Hadoop InteractiVE est une librairie récente (02-2014), dont il faut suivre le développement, évaluer les facilités, performances. Elle pourrait se substituer aux services de Rhadoop listé ci-dessous.
	\item[\tt segue] est dédiée à l'utilisation des services d'Amazon (AWS) à partir de R. 
\end{description}
\subsubsection*{RHadoop}
\href{http://www.revolutionanalytics.com/}{Revolution Analytics} est une
entreprise commerciale qui propose autour de R du support, des services, des
modules d'interface (SAS, SPSS, Teradata, Web, AWS,...) ou d'environnement de
développement sous Windows, de calcul intensif. Comme produit d'appel, elle
soutient le développement de
\href{https://github.com/RevolutionAnalytics/RHadoop/wiki}{Rhadoop}, ensemble de
librairies R d'interface avec \emph{Hadoop}. Cet ensemble de librairies,
développés sous licence libre, comprend :

\centerline{\includegraphics[width=4cm]{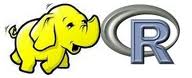}}

\begin{description}
	\item[\href{https://github.com/RevolutionAnalytics/rhdfs}{\tt rhdfs}]   pour utiliser les commandes HDFS d'interrogation d'une base \emph{Hadoop}.
	\item[\href{https://github.com/RevolutionAnalytics/rhbase}{\tt rhbase}] pour utiliser les commandes HBase d'interrogation.
	\item[\href{https://github.com/RevolutionAnalytics/rmr2}{\tt rmr2}] pour exécuter du \emph{MapReduce}; une option permet de tester des scripts (simuler un cluster) sans \emph{hadoop}.
	\item[\href{https://github.com/RevolutionAnalytics/plyrmr}{\tt plyrmr}] utilise la précédente et introduit des facilités.
\end{description}
Certains de ces librairies ne fonctionnent qu'avec \emph{Hadoop} installé
(contrairement à \texttt{rmr2} qui dispose de l'option \texttt{backend="local"}
pour fonctionner localement, c'est à dire en utilisant la mémoire vive comme si
il s'agissait d'un système de fichier HDFS) et sont donc dépendantes du package
\texttt{rJava}.
Quelques
\href{https://github.com/RevolutionAnalytics/RHadoop/wiki/Learning-resources}{
ressources pédagogiques} sont disponibles notamment pour installer RHadoop sur
une machine virtuelle \emph{cloudera}. 

\emph{Revolution Analytics} promeut les librairies de RHadoop avec l'argument recevable que cette solution conduit à des programmes plus courts, comparativement à {\tt HadoopStreaming, RHipe, hive}, donc plus simples et plus rapides à mettre en place mais pas nécessairement plus efficaces en temps de calcul. Une expérimentation en vraie grandeur réalisée par EDF\footnote{Leeley D. P. dos Santos, Alzennyr G. da Silva, Bruno Jacquin, Marie-Luce Picard, David Worms,Charles Bernard (2012). Massive Smart Meter Data Storage and Processing on top of Hadoop. \emph{Workshop Big Data}, Conférence VLDB (Very Large Data Bases), Istambul, Turquie.} montre des exécutions dix fois plus longues de l'algorithme $k$-means avec \emph{RHadoop} qu'avec la librairie \emph{Mahout}. 

\subsubsection*{SparkR} 
Le site collaboratif de
\href{https://github.com/amplab-extras/SparkR-pkg}{SparkR} fournit les premiers
éléments pour installer cette librairie. Celle-ci étant très récente (janvier
2014), il est difficile de se faire un point de vue sans expérimentation en
vraie grandeur. Son installation nécessite celle préalable de \emph{Scala},
\emph{Hadoop} et de la librairie {\tt rJava}.

\section{Algorithmes}
Le développement de cette section suit nécessairement celui des outils disponibles. Elle sera complétée dans les versions à venir de cet article et est accompagnée par un \href{http://wikistat.fr/pdf/st-tutor5-R-mapreduce.pdf}{tuteuriel} réalisé avec \emph{RHadoop} pour aider le statisticien à s'initier aux arcanes de \emph{MapReduce}.
\subsection{Choix en présence}
Après avoir considéré le point de vue ``logiciel'', cette section s'intéresse aux méthodes programmées ou programmables dans un environnement de données massives géré principalement par \emph{Hadoop}. Certains choix sont fait \emph{a priori}:
\begin{itemize}
	\item Si un autre SGBD est utilisé, par exemple \href{https://www.mysql.fr/}{MySQL}, celui-ci est très généralement interfacé avec R ou un langage de requête permet d'en extraire les données utiles. Autrement dit, mis à part les questions de parallélisation spécifiques à \emph{Hadoop} (MapReduce), les autres aspects ne posent pas de problèmes.
	\item Les livres et publications consacrées à \emph{Hadoop} détaillent surtout
des objectifs élémentaires de dénombrement. Comme déjà écrit, ceux-ci ne
nécessitent pas de calculs complexes et donc de programmation qui justifient
l'emploi de R. Nous nous focalisons sur les méthodes dites d'apprentissage
statistique, supervisées ou non.
\end{itemize}
La principale question est de savoir comment un algorithme s'articule avec les fonctionnalités \emph{MapReduce} afin de passer à l'échelle du volume. La contrainte est forte, ce passage à l'échelle n'est pas toujours simple ou efficace et en  conséquence l'architecture \emph{Hadoop} induit une sélection brutale parmi les très nombreux algorithmes d'apprentissage. Se trouvent principalement décrits dans 
\begin{itemize}
\item \emph{Mahout}: système de recommandation, panier de la ménagère, SVD, $k$-means, régression logistique, classifieur bayésien naïf, random forest, SVM (séquentiel),  
\item \emph{RHadoop}: $k$-means,  régression, régression logistique, random forest.
\item \emph{MLbase} de \emph{Spark}: $k$-means, régression linéaire et logistique, système de recommandation, classifieur bayésien naïf et à venir: NMF, CART, random forest.
\end{itemize}

Trois points sont à prendre en compte: la rareté des méthodes d'apprentissage
dans les librairies et même l'absence de certaines méthodes très connues
comparativement à celles utilisables en R, les grandes difficultés quelques fois
évoquées mais pas résolues concernant les réglages des paramètres par validation
croisée ou bootstrap (complexité des modèles), la possibilité d'estimer un
modèle sur un échantillon représentatif. Toutes ces raisons rendent
\emph{indispensable} la disponibilité d'une procédure d'échantillonnage simple
ou équilibré pour anticiper les problèmes.

La plupart des nombreux documents disponibles insistent beaucoup sur
l'installation et l'implémentation des outils, leur programmation, moins sur
leurs propriétés,  ``statistiques''. \emph{Mahout} est développé depuis plus
longtemps et les applications présentées, les stratégies développées, sont très
fouillées pour des volumes importants de données. En revanche, programmés en
java et très peu documentés, il est difficile de rentrer dans les algorithmes
pour apprécier les choix réalisés. Programmés en R les algorithmes de
\emph{Rhadoop} sont plus abordables pour un béotien qui cherche à s'initier.

\subsection{Jeux de données}
Cette section est à compléter notamment avec des jeux de données hexagonaux. La politique de l'état et des collectivités locales d'ouvrir les \href{http://www.data.gouv.fr/}{accès aux données publiques} peut y contribuer de même que le site de concours
``\href{http://datascience.net/fr/home/}{data science}'' mais rien n'est moins
sûr car la CNIL veille, à juste raison, pour rendre impossible les croisements
de fichiers dans l'attente d'une anonymisation efficace (Bras; 2013)\cite{bra13}. Chacun de ceux-ci peut comporter beaucoup de lignes pour finalement très peu de variables; il sera difficile de trouver et extraire des fichiers consistants.

Par ailleurs des  sites proposent des jeux de données pour tester et comparer des méthodes d'apprentissage : \href{http://archive.ics.uci.edu/ml/}{UCI.edu}, \href{http://mldata.org/}{mldata.org}, mais qui, pour l'essentiel, n'atteignent pas les caractéristiques (volume) d'une grande datamasse. D'autres sont spécifiques de données volumineuses.
\begin{itemize}
	\item \href{http://mahout.apache.org/users/basics/collections.html}{mahout.apache.org} proposent un catalogue de ressources dont beaucoup sont textuelles,
	\item  \href{https://www.kaggle.com/competitions}{kaggle.com} propose régulièrement des concours,
	\item  \href{http://www.kdd.org}{kdd.org} organise la compétition la plus en vue.
	\item \href{http://hadoopilluminated.com/hadoop_book/Public_Bigdata_Sets.html}{hadoopilluminated.com} est un livre en ligne pointant des ensembles de données publiques ou des sites ouverts à de telles données.
	\item ...
	\end{itemize}
		Voici quelques  jeux de données utilisés dans la littérature.
		\begin{itemize}
		\item La meilleure prévision (cité par Prajapati; 2013)\cite{pra13} d'enchères de bulldozers disponibles sur le site \href{http://www.kaggle.com/c/bluebook-for-bulldozers}{kaggle} utilise les forêts aléatoires.
		\item \href{ftp://ftp.cdc.gov/pub/Health_Statistics/NCHS/Datasets/DVS/mortality/mort2009us.zip}{Données publiques} de causes de mortalité en 2009 aux USA cité par Adler (2010)\cite{adl10},
	\item \href{http://en.wikipedia.org/wiki/Wikipedia:Statistics}{Statistiques} des consultations de \href{http://wikipedia.fr/index.php}{Wikipédia}. Les \href{http://dumps.wikimedia.org/other/pagecounts-raw/}{données brutes} sont librement accessible et déjà prétraitées par le site \href{http://stats.grok.se/}{stats.grok.se} que  McIver et Brownstein (2014)\cite{mcib14} utilisent pour concurrencer Google dans la \href{http://www.google.org/flutrends/fr/#FR}{prévision des épidémies de grippe} mais sans  précision de localisation.
	\item à compléter...
\end{itemize}

%http://www.alambic-avenir.org/projects/big-data-et-sante-opportunites/

\subsection{MapReduce pour les nuls}\label{mapreduce-section}
\subsubsection*{Principe général}

Lors de la parallélisation ``standard'' d'un calcul, chacun des n\oe uds
impliqués dans le calcul a accès de manière simple à l'intégralité des données
et c'est uniquement le calcul lui-même qui est distribué (c'est-à-dire découpé)
entre les diverses unités de calcul. La différence lors d'une parallélisation
``MapReduce'' est que les données sont elles-mêmes distribuées sur des n\oe uds
différents. Le coût d'accession des données est donc prohibitif et le principe
de \emph{MapReduce} consiste essentiellement à imposer qu'un n\oe ud, dans la phase
``map'', ne travaille que sur des données stockées dans un seul cluster pour
produire en sortie un ordonancement des données qui permet de les répartir de
manière efficace sur des n\oe uds différents pour la phase de calcul elle-même
(phase ``reduce'').

\begin{figure}
\centerline{\includegraphics[width=11.5cm]{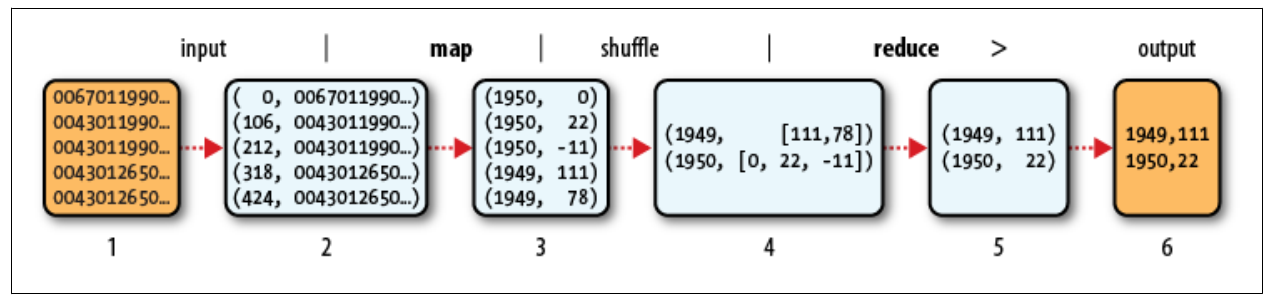}}
\caption{\it Flots des données dans les étapes \emph{MapReduce} (Adler; 2010)\cite{adl10}.}\label{mapreduce}
\end{figure}

Les cinq étapes de parallélisation des calculs se déclinent selon le schéma
ci-dessous. Selon l'algorithme concerné et l'architecture (nombre de serveurs)
utilisée, toutes ne sont pas nécessairement exécutées tandis que certaines
méthodes  nécessitent des itérations du processus jusqu'à convergence. Les
fonctions \emph{Map} et \emph{Reduce} sont écrites dans un quelconque langage,
par exemple R.
\begin{enumerate}
	\item Préparer l'entrée de la phase \emph{Map}. Chaque serveur lit sa part de la base de données ligne par ligne et construit les paires (clef, valeur) pour produire {\tt Map input: list (k1, v1)}.
	\item Exécuter le programme utilisateur de la fonction {\tt Map()} pour émettre {\tt Map output: list (k2, v2)}.
	\item Répartir (\emph{shuffle, combine} ou tri) la liste précédente vers les n\oe uds réducteurs en groupant les clefs similaires comme entrée d'un même n\oe ud. Production de {\tt Reduce input: (k2, list(v2))}. Il n'y a pas de code à fournir pour cette étape qui est implicitement prise en charge.
	\item Pour chaque clef ou chaque pile, un serveur exécute le programme utilisateur de la fonction {\tt Reduce()}. Émission de {\tt Reduce output: (k3, v3)}.
	\item Le n\oe ud maître collecte les sorties et produit, enregistre dans la base, les résultats finaux.
\end{enumerate}
Adler (2010)\cite{adl10} schématise avec la figure \ref{mapreduce} le flot des données. 

\subsubsection*{Exemple trivial}
Considérons un fichier fictif de format texte où les champs, qui décrivent des appels téléphoniques, sont séparés par des virgules et contiennent, entre autres, les informations suivantes. 

{\small \verb+{date},{numéro appelant},...,{numéro appelé},...,{durée}+}
\begin{description}
	\item[\tt Map] ~
		\begin{itemize}
		\item reçoit une ligne ou enregistrement du fichier, 
		\item utilise un langage de programmation pour extraire la date et la durée, 
		\item émet une paire : clef (date) et valeur (durée).
	\end{itemize}
	\item[\tt Reduce] ~
	\begin{itemize}
	\item reçoit une paire: clef (date) et valeur (durée$_1$ ... durée$_n$), 
	\item programme une boucle qui calcule la somme des durées et le nombre d'appels,
	\item calcule la moyenne, 
	\item sort le résultat: (date) et (moyenne, nombre d'appels).
	\end{itemize}
\end{description}

Si l'objectif est maintenant de compter le nombre d'appels par jour et par numéro d'appel, la clef de l'étape {\tt Map} se complique tandis que l'étape {\tt Reduce} se réduit à un simple comptage.

\begin{description}
	\item[\tt Map] ~
		\begin{itemize}
		\item reçoit une ligne ou enregistrement du fichier, 
		\item extrait la date et le numéro d'appel, 
		\item émet une paire : clef (date, numéro) et valeur (1).
	\end{itemize}
		\item[\tt Reduce] ~
		\begin{itemize}
	\item reçoit une clef: (date, numéro) et valeur (1...1), 
	\item boucle qui calcule la somme des valeurs égale au nombre d'appels,
	\item  sort le résultat: (date, numéro) et (nombre).
	\end{itemize}
\end{description}
Toute l'astuce réside donc dans la façon de gérer les paires (clef, valeur) qui définissent les échanges entre les étapes. Ces paires peuvent contenir des objets complexes (vecteurs, matrices, listes). 

\subsection{Échantillonnage aléatoire simple}
Des outils de base sont indispensables pour extraire un échantillon d'une base \emph{Hadoop} et revenir à un environnement de travail classique pour disposer des outils de modélisation, choix  et optimisation des modèles.

\subsubsection*{\emph{Reservoir Sampling}}
L'objectif est d'extraire, d'une base de taille $N$ (pas nécessairement connu), un échantillon représentatif de taille $n$ par tirage aléatoire simple en s'assurant que chaque observation ait la même probabilité $n/N$ d'être retenue. Attention, la répartition des observations selon les serveurs peut-être biaisée. La contrainte est celle d'\emph{Hadoop}, pas de communication entre les serveurs, et il faut minimiser le temps donc se limiter à une seule lecture de l'ensemble de la base. 

L'algorithme dit de \emph{reservoir sampling} (Vitter; 1985) tient ces objectifs si l'échantillon retenu tient en mémoire dans un seul n\oe ud. Malheureusement cette méthode est par principe séquentielle et les aménagements pour la paralléliser avec \emph{MapReduce} peut poser des problèmes de représentativité de l'échantillon, notamment si les capacités de stockage des n\oe uds sont déséquilibrées.

\begin{algorithmic}[1]
	\REQUIRE {$\bm{s}$ : ligne courante de la base (lue ligne à ligne)}
	\REQUIRE{$\bm{R}$ : matrice réservoir de $n$ lignes}
	\FOR{$i=1$ \TO $n$}
		\STATE Retenir les $n$ premières lignes : $\bm{R}_i\gets \bm{s}(i)$
	\ENDFOR
	\STATE $i \gets n+1$
	\REPEAT
		\STATE Tirer un nombre entier $j$ aléatoirement entre 1 et $i$ (inclus)
		\IF{$j \leq n$}
			\STATE $\bm{R}_i \gets \bm{s}(i)$
		\ENDIF
		\STATE $i \gets i+1$
	\UNTIL{toutes les observations ont été lues ($\Leftrightarrow i$ est la taille
de $\bm{s}$)} 
\end{algorithmic}

La phase Map est triviale tandis que celle Reduce, tire un entier aléatoire et
assure le remplacement conditionnel dans la partie ``valeur'' de la sortie. Les
clefs sont sans importance. Vitter (1985)\cite{vit85} montre que cet algorithme
conçu pour échantillonner sur une bande magnétique assure le tirage avec
équiprobabilité. Des variantes: pondérées, équilibrées, stratifiées, existent
mais celle échelonnable pour \emph{Hadoop} pose des problèmes si les donnéees
sont mal réparties sur les n\oe uds.

% http://had00b.blogspot.fr/2013/07/random-subset-in-mapreduce.html
\subsubsection*{Par tri de l'échantillon}
Le principe de l'algorithme est très élémentaire; $N$ nombres aléatoires uniformes sur $[0,1]$ sont tirés et associés: une clef à chaque observation ou valeur. Les paires (clef, observation) sont triées selon cette clef et les $n$ premières ou $n$ dernières sont retenues. 

L'étape \emph{Map} est facilement définie de même que l'étape \emph{Reduce} de sélection mais l'étape intermédiaire de tri peut être très lourde car elle porte sur toute la base.

\subsubsection*{ScanSRS}
Xiangrui (2013)\cite{xia13} propose de mixer les deux principes en sélectionnant un sous-ensemble des observations les plus raisonnablement probables pour réduire le volume de tri. L'inégalité de Bernstein est utilisée pour retenir ou non des observations dans un réservoir de taille en $0(n)$ car nécessairement plus grand que dans l'exemple précédent; observations qui sont ensuite  triées sur leur clef avant de ne conserver que les $n$ plus petites clefs. Un paramètre règle le risque de ne pas obtenir au moins $n$ observations tout en contrôlant la taille du réservoir.

\subsection{Factorisation d'une matrice}
Tous les domaines de datamasse produisent, entre autres, de très grandes
matrices creuses: clients réalisant des achats, notant des films, relevés de
capteurs sur des avions, voitures, textes et occurrences de mots... objets avec
des capteurs, des gènes avec des expressions. Alors  que ces méthodes sont
beaucoup utilisés en e-commerce, les acteurs se montrent discrets sur leurs
usages et les codes exécutés, de même que la fondation Apache sur la NMF.
Heureusement, la recherche en biologie sur fonds publics permet d'y remédier.

\subsubsection*{SVD}
Les algorithmes de décomposition en valeurs singulières
(\href{http://wikistat.fr/pdf/st-m-explo-alglin.pdf}{SVD}) d'une grande matrice
creuse,  sont connus de longue date mais exécutés sur des machines massivement
parallèles de calcul intensif pour la résolution de grands systèmes linéaires en
analyse numérique. Leur adaptation au cadre \emph{MapReduce} et leur
application à l'\href{http://wikistat.fr/pdf/st-m-explo-acp.pdf}{analyse en
composantes principales} ou
l'\href{http://wikistat.fr/pdf/st-m-explo-afc.pdf}{analyse des correspondances},
ne peut se faire de la même façon et conduit à d'autres algorithmes. 

\href{https://mahout.apache.org/users/dim-reduction/dimensional-reduction.html}{Mahout} propose deux implémentations \emph{MapReduce} de la SVD et une autre version basée sur un algorithme stochastique. En juin 2014, \emph{RHadoop} ne propose pas encore d'algorithme, ni \emph{Spark}.

\subsubsection*{NMF}
La factorisation de matrices non négatives (Paatero et Tapper;
1994\cite{paat94}, Lee et Seung; 1999\cite{lees99}) est plus récente. Sûrement
très utilisée par les entreprises commerciales, elle n'est cependant pas
implémentée dans les librairies ouvertes (\emph{RHadoop, Mahout}) basées sur
\emph{Hadoop}. Il existe une librairie R ({\tt NMF}) qui exécute au choix
plusieurs algorithmes selon deux critères possibles de la factorisation non
négative et utilisant les fonctionnalités de parallélisation d'une machine
(multi c\oe urs) grâce à la librairie {\tt parallel} de R. Elle n'est pas
interfacée avec \emph{Hadoop} alors que celle plus rudimentaire  développée en
java par Ruiqi et al. (2014)\cite{ruiyj14} est associée à \emph{Hadoop}. 

\subsection{$k$-means}
L'algorithme $k$-means et d'autres de classification non supervisée sont adaptés à \emph{Hadoop} dans la plupart des librairies. Le principe des algorithmes utilisés consiste à itérer un nombre de fois fixé \emph{a priori} ou jusqu'à convergence les étapes \emph{Map-Reduce}. Le principal problème lors de l'exécution est que chaque itération provoque une réécriture des données pour ensuite les relire pour l'itération suivante. C'est une contrainte imposée par les fonctionnalités \emph{MapReduce} de \emph{Hadoop} (à l'exception de \emph{Spark}) pour toute exécution d'un algorithme itératif qui évidemment pénalise fortement le temps de calcul. 

Bien entendu, le choix de la fonction qui calcule la distance entre une matrice de $k$ centres et une matrice d'individus est fondamental.

	\begin{enumerate}
		\item L'étape Map utilise cette fonction pour calculer un paquet de distances et retourner le centre le plus proche de chaque individu. Les individus sont bien stockés dans la base HDFS alors que les centres, initialisés aléatoirement restent en mémoire.
		\item Pour chaque clef désignant un groupe, l'étape Reduce calcule les nouveaux barycentres, moyennes des colonnes des individus partageant la même classe / clef.
		\item Un programme global exécute une boucle qui itère les étapes
\emph{MapReduce} en lisant / écrivant (sauf pour \emph{Spark}) à chaque 
itération les individus dans la base et remettant à jour les centres.
	\end{enumerate}

\subsection{Régression linéaire}

Est-il pertinent ou réellement utile, en terme de qualité de prévision, d'estimer un \href{http://wikistat.fr/pdf/st-m-modlin-regmult.pdf}{modèle de régression} sur un échantillon de très grande taille alors que les principales difficultés sont généralement  soulevées par les questions  de sélection de variables, sélection de modèle. De plus, les fonctionnalités offertes sont très limitées, sans aucune aide au diagnostic. Néanmoins, la façon d'estimer un modèle de régression illustre une autre façon d'utiliser les fonctionnalités \emph{MapReduce}  pour calculer notamment des produits matriciels.

Soit $\bm{X}$ la matrice $(n\times (p+1))$ (\emph{design matrix}) contenant en
première colonne des 1 et les observations des $p$ variables explicatives
quantitatives sur les $n$ observations; $\bm{y}$ le vecteur des observations de
la variable à expliquer.

On considère $n$ trop grand pour que la matrice $\bm{X}$ soit chargée en mémoire mais $p$ pas trop grand pour que le produit $\bm{X'X}$ le soit. 

Pour estimer le modèle 
$$\bm{y}=\bm{X}\bs{\beta}+\bs{\varepsilon},$$
l'algorithme combine deux étapes de \emph{MapReduce} afin de calculer chaque produit matriciel $\bm{X'X}$ et $\bm{X'y}$. Ensuite, un appel à la fonction {\tt solve} de R résout les équations normales 
$$\bm{X'X} = \bm{X'y}$$
et fournit les estimations des paramètres $\beta_j$. 

Le produit matriciel est décomposé en la somme (étape Reduce) de la liste des
matrices issues des produits (étape Map) $\bm{X}_i'\bm{X}_j (i,j = 1,...,p)$ 
(ces produits sont calculés pour des sous-ensembles petits de données puis 
sommés).

\subsection{Régression logistique}
Les mêmes réserves que ci-dessus sont faites pour la \href{http://wikistat.fr/pdf/st-m-app-rlogit.pdf}{régression logistique} qui est obtenue par un algorithme de descente du gradient arrêté après un nombre fixe d'itérations et dont chacune est une étape \emph{MapReduce}.  L'étape \emph{Map} calcule la contribution de chaque individu au gradient puis l'écrit dans la base, l'étape \emph{Reduce} est une somme pondérée. Chaque itération provoque donc $n$ lectures et $n$ écritures dans la base. 

Des améliorations pourraient être apportées: adopter un critère de convergence pour arrêter l'algorithme plutôt que fixer le nombre d'itérations, utilisé un gradient conjugué. La faiblesse de l'algorithme reste le nombre d'entrées/sorties à chaque itération. 

\subsection{Random forest}
\href{http://wikistat.fr/pdf/st-m-app-agreg.pdf}{Random Forest}, cas particulier de \href{http://wikistat.fr/pdf/st-m-app-agreg.pdf}{bagging}, s'adapte particulièrement bien à l'environnement d'\emph{Hadoop} lorsqu'une grande taille de l'ensemble d'apprentissage fournit des échantillons indépendants pour estimer et agréger (moyenner) des ensembles de modèles. Comme en plus, \emph{random forest} semble insensible au sur-apprentissage, cette méthode ne nécessite généralement pas de gros efforts d'optimisation de paramètres, elle évite donc l'un des principaux écueils des approches Big Data en apprentissage. 

\subsubsection*{Principe}
Soit $n$ la taille totale (grande) de l'échantillon d'apprentissage, $m$ le
nombre d'arbres ou modèles de la forêt et $k$ la taille de chaque échantillon
sur lequel est estimé un modèle. Les paramètres $m$ et $k$, ainsi que {\tt mtry}
(le nombre de variables aléatoirement sélectionnées parmi les variables
initiales pour définir un n\oe ud dans chacun des arbres) sont en principe
à optimiser mais si $n$ est grand, on peut espérer qu'ils auront peu
d'influence. Il y a en principe trois cas à considérer en fonction de la taille
de $n$. 
\begin{itemize}
\item $k\times m < n$ toutes les données ne sont pas utilisées et des
échantillons indépendants sont obtenus par tirage aléatoire sans remise, 
\item $k\times m = n$ des échantillons indépendants sont exactement obtenus par
tirages sans remise, 
\item $k\times m > n$ correspond à une situation où un ré-échantillonnage avec remise est nécessaire ; $k=n$ correspond à la situation classique du \href{http://wikistat.fr/pdf/st-m-app-bootstrap.pdf}{bootstrap}.
\end{itemize}
L'astuce de \href{http://blog.cloudera.com/blog/2013/02/how-to-resample-from-a-large-data-set-in-parallel-with-r-on-hadoop/}{Uri Laserson} est de traiter dans un même cadre les trois situations en considérant des tirages selon des lois de Poisson pour approcher celui multinomial correspondant au bootstrap. L'idée consiste donc à tirer indépendamment pour chaque observation selon une loi de Poisson.

\subsubsection*{Échantillonnage}
L'objectif est d'éviter plusieurs lectures, $m$, de l'ensemble des données pour réaliser $m$ tirages avec remise et également le tirage selon une multinomiale qui nécessite des échanges de messages, impossible dans \emph{Hadoop}, entre serveurs. 

Le principe de l'approximation est le suivant: pour chaque observation $x_i
(i=1,..,n)$  tirer $m$ fois selon une loi de Poisson de paramètre $(k/n)$, une
valeur $p_{i,j}$ pour chaque modèle $M_j (j=1,...,m)$ (correspondant à l'arbre
$j$ dans le cadre des forêts aléatoires) qui correspond à l'arbre $j$. L'étape
Map est ainsi constituée : elle émet un total de $p_{i,j}$ fois la paire (clef,
valeur) $(j, x_i)$; $p_{i,j}$ pouvant être 0. Même si les observations ne sont
pas aléatoirement réparties selon les serveurs, ce tirage garantit l'obtention
de $m$ échantillons aléatoires dont la taille est approximativement $k$. D'autre
part, approximativement $\exp(-km/n)$ des données initiales ne seront présentes
dans aucun des échantillons. 

L'étape de tri ou \emph{shuffle} redistribue les échantillons identifiés par
leur clef à chaque serveur qui estime un arbres (ou plusieurs successivement).
Ces arbres sont stockés pour constituer la forêt nécessaire à la prévision. 

Cette approche, citée par Prajapati (2010)\cite{pra13}, est testée sur les enchères de bulldozer de \href{http://www.kaggle.com/c/bluebook-for-bulldozers}{Kaggle} par \href{http://blog.cloudera.com/blog/2013/02/how-to-resample-from-a-large-data-set-in-parallel-with-r-on-hadoop/}{Uri Laserson} qui en développe le code.

\section*{Conclusion}

Quelques remarques pour conclure ce tour d'horizon des outils permettant d'exécuter des méthodes d'apprentissage supervisé ou non sur des données volumineuses. 
\begin{itemize}
	\item Cet aperçu est un instantané à une date donnée, qui va évoluer rapidement en fonction des développements du secteur et donc des environnements, notamment ceux \emph{Mahout} et \emph{Spark} à suivre attentivement. C'est l'avantage d'un support de cours électronique car adaptable en temps réel. 
	\item L'apport pédagogique important et difficile concerne la bonne connaissance des méthodes de modélisation, de leurs propriétés, de leurs limites. Pour cet objectif, R reste un outil à privilégier. Largement interfacé avec tous les systèmes de gestion de données massives ou non, il est également utilisé en situation ``industrielle''.  
	\item Une remarque fondamentale concerne le réglage ou l'optimisation des paramètres des méthodes d'apprentissage: nombre de classes, sélection de variables, paramètre de complexité. Ces problèmes, qui sont de réelles difficultés pour les méthodes en question et qui occupent beaucoup les équipes travaillant en apprentissage (machine et/ou statistique), sont largement passés sous silence. Que deviennent les procédures de validation croisée, \emph{bootstrap}, les échantillons de validation et test, bref toute la réflexion sur l'optimisation des modèles pour s'assurer de leur précision et donc de leur validité?

	\item Si R, éventuellement précédé d'une phase d'échantillonnage (C, java, python, perl) n'est plus adapté au volume et
	\item Si \emph{RHadoop} s'avère trop lent, 
	\item \emph{Mahout} et \emph{Spark} sont à tester car semble-t-il nettement plus efficaces. Un apprentissage élémentaire de java, s'avère donc fort utile pour répondre à cet objectif à moins que la mise en exploitation du langage matriciel progresse rapidement.
	\item Si ce n'est toujours pas suffisant, notamment pour prendre en compte les composantes \href{}{Variété} et \href{}{Vélocité} du \emph{Big Data}, les compétences acquises permettent au moins la réalisation de prototypes avant de réaliser ou faire réaliser des développements logiciels plus conséquents. Néanmoins le passage à l'échelle est plus probant si les prototypes anticipent la parallélisation en utilisant les bons langages comme \href{http://www.scala-lang.org/}{scala} ou \href{http://clojure.org/}{clojure}. 
\end{itemize}

Ce rapide descriptif technique des outils d'analyse de la datamasse n'aborde pas les très nombreuses implications sociétales largement traitées par les médias ``grand public'' qui alimente une large confusion dans les objectifs poursuivis, par exemple  entre les deux ci-dessous. 
\begin{itemize}
	\item Le premier est d'ordre ``clinique'' ou d'``enquête de police''. Si une
information est présente dans les données, sur le web, les algorithmes
d'exploration exhaustifs vont la trouver; c'est la NSA dans la \emph{La Menace
Fantôme}.
	\item Le deuxième est ``prédictif''. Apprendre des comportements, des occurrences d'évènements pour \emph{prévoir} ceux à venir; c'est GAFA dans \emph{L'Empire Contre Attaque}. Il faut ensuite distinguer entre :
	\begin{itemize}
		\item prévoir un comportement moyen, une tendance moyenne, comme celle souvent citée de l'évolution d'une \href{http://www.google.org/flutrends/intl/fr/fr/#FR}{épidémie de grippe} ou encore de ventes de jeux vidéos, chansons, films... (Goel et al., 2010)\cite{goehl10} à partir des fréquences des mots clefs associés recherchés sur \emph{Google} et
		\item prévoir un comportement individuel, si M. Martin habitant telle adresse va ou non attraper la grippe, acheter tel film!
	\end{itemize}
\end{itemize}
Il est naïf de penser que l'efficacité obtenue pour l'atteinte du premier
objectif peut se reporter sur le deuxième pour la seule raison que ``toutes''
les données sont traitées. La fouille de données (\emph{data mining}), et depuis
plus longtemps la Statistique, poursuivent l'objectif de prévision avec des taux de réussite, ou d'erreur, qui atteignent nécessairement des limites que ce soit avec 1000, 10 000 ou 10 millions... d'observations. L'estimation d'une moyenne, d'une variance, du taux d'erreur, s'affinent et convergent vers la ``vraie'' valeur (loi des grands nombres), mais l'erreur d'une \emph{prévision individuelle}, et son taux reste inhérents à une variabilité intrinsèque et irréductible, comme celle du vivant. Le \emph{Retour du Jedi} est, ou sera, celui du hasard ou celui du \emph{libre arbitre}. 

\setlength{\fboxrule}{0pt}
\centerline{\fbox{%
\begin{minipage}[c]{0.25\textwidth}
\centerline{\includegraphics[width=2.5cm]{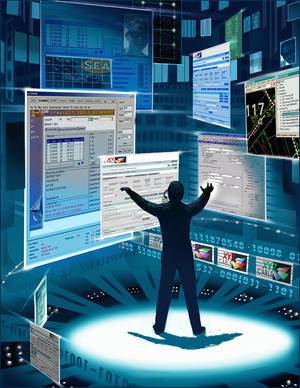}}
\end{minipage}
\begin{minipage}[c]{0.25\textwidth}
En résumé: \emph{Back to the Futur} et \emph{Star War}, mais pas \emph{Minority Report}.
\end{minipage}%
}}

En revanche et de façon positive,  l'accumulation de données d'origines
différentes, sous contrôle de la CNIL et sous réserve d'une anonymisation
rigoureuse empêchant une ré-identification (Bras; 2013)\cite{bra13}, permet
d'étendre le deuxième objectif de prévision à bien d'autres champs d'application
que ceux commerciaux; par exemple en santé publique, si le choix politique en
est fait et les données publiques efficacement protégées.

Le domaine de l'assurance\footnote{Lire à ce propos un \href{http://www.revue-banque.fr/banque-detail-assurance/article/big-data-defis-opportunites-pour-les-assureurs}{résumé des interventions} (P. Thourot et F. Ewald, \emph{Revue Banque}, N°315, juin 2013) ou les \href{http://www.enass.fr/recherche/conferences/conferences-de-l-ecole-nationale-d-assurances-fr-5-c28.html}{actes} d'un colloque: Big Data Défis et Opportunités pour les Assureurs.} illustre bien les enjeux, bénéfices et risques potentiels associés au traitement des données massives. La fouille de données est déjà largement utilisée depuis la fin du siècle dernier pour cibler la communication aux clients mais comment la réglementation va ou doit évoluer si ces masses d'informations (habitudes alimentaires, localisation géographiques, génome, podomètre et usages sportifs, réseaux sociaux...) deviennent utilisables pour segmenter et adapter la tarification des contrats, autos, vie, complémentaires de santé ou lors de prêts immobiliers. L'
assurance ou la mutualisation des risques repose sur un principe d'\emph{asymétrie d'information}. Si celle-ci est inversée au profit de l'assureur, par exemple si GAFA propose des contrats, ce serait à l'encontre de toute idée de mutualisation des risques.

\bibliographystyle{babamspl}
\bibliography{biblio_bigdata}

\begin{thebibliography}{10}
  \providecommand{\bysame}{\leavevmode\hbox to3em{\hrulefill}\thinspace}
  \providecommand{\MR}{\relax\ifhmode\unskip\space\fi MR }
  % \MRhref is called by the amsart/book/proc definition of \MR.
  \providecommand{\MRhref}[2]{%
    \href{http://www.ams.org/mathscinet-getitem?mr=#1}{#2}
  }
  \providecommand{\href}[2]{#2}
  \providebibliographyfont{name}{}%
  \providebibliographyfont{lastname}{}%
  \providebibliographyfont{title}{\emph}%
  \providebibliographyfont{jtitle}{\btxtitlefont}%
  \providebibliographyfont{etal}{}%
  \providebibliographyfont{journal}{}%
  \providebibliographyfont{volume}{\textbf}%
  \providebibliographyfont{ISBN}{\MakeUppercase}%
  \providebibliographyfont{ISSN}{\MakeUppercase}%
  \providebibliographyfont{url}{\url}%
  \providebibliographyfont{numeral}{}%
  \providecommand\btxprintamslanguage[1]{\ (#1)}
  \expandafter\btxselectlanguage\expandafter {\btxfallbacklanguage}

\expandafter\btxselectlanguage\expandafter {\btxfallbacklanguage}
\bibitem {adl10}
\btxnamefont {Joseph \btxlastnamefont {Adler}}, \btxtitlefont {\btxifchangecase
  {R in a nutshell}{R in a nutshell}}, \btxpublisherfont {O'Reilly}, 2010.

\bibitem {besgl14}
\btxnamefont {Philippe \btxlastnamefont {Besse}}, \btxnamefont {Aur{\'e}lien
  \btxlastnamefont {Garivier}}\btxandcomma {} \btxandlong {} \btxnamefont
  {Jean\btxfnamespacelong Michel \btxlastnamefont {Loubes}}, \btxtitlefont
  {\btxifchangecase {{Big Data Analytics - Retour vers le Futur 3; De
  Statisticien {\`a} Data Scientist}}{{Big Data Analytics - Retour vers le
  Futur 3; De Statisticien {\`a} Data Scientist}}}, {\latintext
  \btxurlfont{http://hal.archives-ouvertes.fr/hal-00959267}}, 2014.

\bibitem {bra13}
\btxnamefont {Pierre\btxfnamespacelong Louis \btxlastnamefont {Bras}},
  \btxtitlefont {\btxifchangecase {Rapport sur la gouvernance et l'utilisation
  des données de santé}{Rapport sur la gouvernance et l'utilisation des données
  de santé}}, {\latintext
  \btxurlfont{http://www.social-sante.gouv.fr/IMG/pdf/Rapport_donnees_de_sante_2013.pdf}},
  2013.

\bibitem {dea04}
\btxnamefont {Jeffrey \btxlastnamefont {Dean}} \btxandlong {} \btxnamefont
  {Sanjay \btxlastnamefont {Ghemawat}}, \btxtitlefont {\btxifchangecase
  {Mapreduce: Simplified data processing on large clusters}{MapReduce:
  Simplified Data Processing on Large Clusters}}, Proceedings of the 6th
  Conference on Symposium on Opearting Systems Design \& Implementation -
  Volume 6, OSDI'04, 2004.

\bibitem {goehl10}
\btxnamefont {Sharad \btxlastnamefont {Goel}}, \btxnamefont
  {Jake\btxfnamespacelong M. \btxlastnamefont {Hofman}}, \btxnamefont
  {Sébastien \btxlastnamefont {Lahaie}}, \btxnamefont {David\btxfnamespacelong
  M. \btxlastnamefont {Pennock}}\btxandcomma {} \btxandlong {} \btxnamefont
  {Duncan\btxfnamespacelong J. \btxlastnamefont {Watts}}, \btxjtitlefont
  {\btxifchangecase {Predicting consumer behavior with web search}{Predicting
  consumer behavior with Web search}}, \btxjournalfont {Proceedings of the
  National Academy of Sciences} (2010).

\bibitem {lees99}
\btxnamefont {D.~\btxlastnamefont {Lee}} \btxandlong {} \btxnamefont
  {S.~\btxlastnamefont {Seung}}, \btxjtitlefont {\btxifchangecase {Learning the
  parts of objects by non-negative matrix factorization}{Learning the parts of
  objects by non-negative matrix factorization}}, \btxjournalfont {Nature}
  (1999).

\bibitem {ruiyj14}
\btxnamefont {Ruiqi \btxlastnamefont {Liao}}, \btxnamefont {Yifan
  \btxlastnamefont {Zhang}}, \btxnamefont {Jihong \btxlastnamefont
  {Guan}}\btxandcomma {} \btxandlong {} \btxnamefont {Shuigeng \btxlastnamefont
  {Zhou}}, \btxjtitlefont {\btxifchangecase {Cloudnmf: A mapreduce
  implementation of nonnegative matrix factorization for large-scale biological
  datasets}{CloudNMF: A MapReduce Implementation of Nonnegative Matrix
  Factorization for Large-scale Biological Datasets}}, \btxjournalfont
  {Genomics, Proteomics \& Bioinformatics} \btxvolumefont {12} (2014),
  \btxnumbershort {.}~1, 48 -- 51.

\bibitem {xia13}
\btxnamefont {Xiangrui \btxlastnamefont {M.}}, \btxtitlefont {\btxifchangecase
  {Scalable simple random sampling and statified sampling}{Scalable Simple
  Random Sampling and Statified Sampling}}, Proceedings of the 30th
  International Conference on Machine Learning, 2013.

\bibitem {mccawl11}
\btxnamefont {E.~\btxlastnamefont {McCallum}} \btxandlong {} \btxnamefont
  {S.~\btxlastnamefont {Weston}}, \btxtitlefont {\btxifchangecase {Parallel
  r}{Parallel R}}, \btxpublisherfont {O'Reilly Media}, 2011.

\bibitem {mcib14}
\btxnamefont {David \btxlastnamefont {McIver}} \btxandlong {} \btxnamefont
  {John \btxlastnamefont {Brownstein}}, \btxjtitlefont {\btxifchangecase
  {Wikipedia usage estimates prevalence of influenza-like illness in the united
  states in near real-time}{Wikipedia Usage Estimates Prevalence of
  Influenza-Like Illness in the United States in Near Real-Time}},
  \btxjournalfont {PLoS Comput Biol} \btxvolumefont {10} (2014),
  \btxnumbershort {.}~4, {\latintext
  \btxurlfont{http://dx.doi.org/10.1371%2Fjournal.pcbi.1003581}}.

\bibitem {oweadf11}
\btxnamefont {Sean \btxlastnamefont {Owen}}, \btxnamefont {Robin
  \btxlastnamefont {Anil}}, \btxnamefont {Ted \btxlastnamefont
  {Dunning}}\btxandcomma {} \btxandlong {} \btxnamefont {Ellen \btxlastnamefont
  {Friedman}}, \btxtitlefont {\btxifchangecase {Mahout in action}{Mahout in
  Action}}, \btxpublisherfont {Manning Publications Co.}, 2011.

\bibitem {paat94}
\btxnamefont {Pentti \btxlastnamefont {Paatero}} \btxandlong {} \btxnamefont
  {Unto \btxlastnamefont {Tapper}}, \btxjtitlefont {\btxifchangecase {Positive
  matrix factorization: A non-negative factor model with optimal utilization of
  error estimates of data values}{Positive matrix factorization: A non-negative
  factor model with optimal utilization of error estimates of data values}},
  \btxjournalfont {Environmetrics} \btxvolumefont {5} (1994), \btxnumbershort
  {.}~2, 111--126.

\bibitem {pra13}
\btxnamefont {Vignesh \btxlastnamefont {Prajapati}}, \btxtitlefont
  {\btxifchangecase {Big data analytics with r and hadoop}{Big Data Analytics
  with R and Hadoop}}, \btxpublisherfont {Packt Publishing}, 2013.

\bibitem {rcor14}
\btxnamefont {\btxlastnamefont {{R Core Team}}}, \btxtitlefont
  {\btxifchangecase {R: A language and environment for statistical
  computing}{R: A Language and Environment for Statistical Computing}}, R
  Foundation for Statistical Computing, 2014, {\latintext
  \btxurlfont{http://www.R-project.org/}}.

\bibitem {vit85}
\btxnamefont {Vitter\btxfnamespacelong J. \btxlastnamefont {S.}},
  \btxjtitlefont {\btxifchangecase {Random sampling with a reservoir}{Random
  sampling with a reservoir}}, \btxjournalfont {ACM transaction on Mathematical
  Software} \btxvolumefont {11:1} (1985), 37--57.

\bibitem {zahc12}
\btxnamefont {Matei \btxlastnamefont {Zaharia}}, \btxnamefont {Mosharaf
  \btxlastnamefont {Chowdhury}}, \btxnamefont {Tathagata \btxlastnamefont
  {Das}}, \btxnamefont {Ankur \btxlastnamefont {Dave}}, \btxnamefont {Justin
  \btxlastnamefont {Ma}}, \btxnamefont {Murphy \btxlastnamefont {McCauly}},
  \btxnamefont {Michael\btxfnamespacelong J. \btxlastnamefont {Franklin}},
  \btxnamefont {Scott \btxlastnamefont {Shenker}}\btxandcomma {} \btxandlong {}
  \btxnamefont {Ion \btxlastnamefont {Stoica}}, \btxtitlefont {\btxifchangecase
  {Resilient distributed datasets: A fault-tolerant abstraction for in-memory
  cluster computing}{Resilient Distributed Datasets: A Fault-Tolerant
  Abstraction for In-Memory Cluster Computing}}, Presented as part of the 9th
  USENIX Symposium on Networked Systems Design and Implementation (NSDI 12)\
  (San Jose, CA), \btxpublisherfont {USENIX}, 2012, \btxpagesshort {.}~15--28,
  \mbox{\btxISBN~\btxISBNfont {978-931971-92-8}}, {\latintext
  \btxurlfont{https://www.usenix.org/conference/nsdi12/technical-sessions/presentation/zaharia}}.

\end{thebibliography}

\end{document}